\documentclass[aps,twocolumn,prl]{revtex4}
\usepackage{graphicx,amssymb,amsmath}
\usepackage{graphicx}
\usepackage{dcolumn}
\usepackage{bm}

\newcommand{\Bpar}{$B_{||}$}
\newcommand{\Bperp}{$B_{\perp}$}
\newcommand{\rxx}{$R_{xx}$}
\newcommand{\ryy}{$R_{yy}$}
\newcommand{\rxy}{$R_{xy}$}
\newcommand{\ryx}{$R_{yx}$}
\newcommand{\ydir}{$\langle 1\overline{1} 0 \rangle$}
\newcommand{\xdir}{$\langle 110 \rangle$}

\begin{document}

\title{Evidence for a fractionally quantized Hall state with anisotropic longitudinal transport}

\author{Jing Xia$^1$, J.P. Eisenstein$^1$, L.N. Pfeiffer$^2$, and K.W. West$^2$}

\affiliation{$^1$Condensed Matter Physics, California Institute of Technology, Pasadena, CA 91125
\\
$^2$Department of Electrical Engineering, Princeton University, Princeton, NJ 08544}

\date{\today}

\begin{abstract} 

\end{abstract}


\maketitle
{\bf
At high magnetic fields, where the Fermi level lies in the $N=0$ lowest Landau level (LL), a clean two-dimensional electron system (2DES) exhibits numerous incompressible liquid phases which display the fractional quantized Hall effect (FQHE) \cite{perspectives}.  These liquid phases do not break rotational symmetry, exhibiting resistivities which are isotropic in the plane.  In contrast, at lower fields, when the Fermi level lies in the $N\ge2$ third and several higher LLs, the 2DES displays a distinctly different class of collective states.  In particular, near half filling of these high LLs the 2DES exhibits a strongly anisotropic longitudinal resistance at low temperatures \cite{Lilly99a,Du99}.  These ``stripe'' phases, which do not exhibit the quantized Hall effect, resemble nematic liquid crystals, possessing broken rotational symmetry and orientational order \cite{KFS,FKS,MC,FK,AR}. Here we report a surprising new observation: An electronic configuration in the $N=1$ second LL whose resistivity tensor {\it simultaneously} displays a robust fractionally quantized Hall plateau and a strongly anisotropic longitudinal resistance resembling that of the stripe phases.
}

The sample (details described in Methods) we employ is a square shaped GaAs/AlGaAs heterostructure with edges parallel to the \xdir\ and \ydir\ crystal directions, henceforth referred to as the $\hat{x}$ and $\hat{y}$ directions, respectively. Substantial in-plane magnetic fields \Bpar\ may be added to the field \Bperp\ perpendicular to the 2DES plane by tilting the sample at low temperatures. For the present studies, \Bpar\ lies along the $\hat{x}$, or \xdir, direction. Figure 1 shows the longitudinal and Hall resistances at $T \approx 15$ mK with the magnetic field perpendicular to the 2DES plane (tilt angle $\theta = 0$).  For the field range shown, the Fermi level lies in the lower spin branch of the $N=1$ LL where the filling fraction $\nu \equiv nh/eB_{\perp}$ runs from $\nu = 2$ to $\nu = 3$. Deep minima in the longitudinal resistances and associated plateaus in \rxy\ and \ryx\ clearly signal the presence of FQHE states at $\nu =7/3$, 5/2, and 8/3. While the $\nu = 7/3 = 2+1/3$ and $8/3 = 2+ 2/3$ states may be kin to the well-known $\nu = 1/3$ and 2/3 FQHE states in the $N=0$ lowest LL \cite{perspectives} (alternatives do exist; see \cite{ReadRezayi}), the $\nu = 5/2$ state \cite{Willett87} is thought to be an example of the non-abelian Moore-Read paired composite fermion state \cite{moore_read}. In addition to these and a few weaker FQHE states, the four known re-entrant integer quantized Hall states \cite{JPE02}, in varying stages of development, are also evident in Fig. 1.  These insulating phases are poorly understood but may be related to the ``bubble'' phases found in the flanks of the $N\ge2$ LLs and which also exhibit re-entrant integer Hall quantization \cite{KFS,FKS,MC,Lilly99a,Du99}.  As the data in Fig. 1 make clear, the longitudinal and Hall resistances are very similar for current flow along \xdir\ and \ydir. (The small differences between \rxx\ and \ryy\ for $2.85 \lesssim B_{\perp} \lesssim 3$ T is likely due to extrinsic sample-dependent effects of no relevance here.) Unlike the situation in the $N\ge2$ LLs, no anisotropic phases have been found in 2D electron systems in the $N=1$ LL, at least in the absence of an external symmetry breaking field such as an in-plane magnetic field \Bpar. (Anisotropy in the $N=1$ LL has been observed in 2D {\it hole} systems \cite{Shayegan} \cite{Manfra}.)

\begin{figure}
\begin{center}
\includegraphics[width=1.0 \columnwidth] {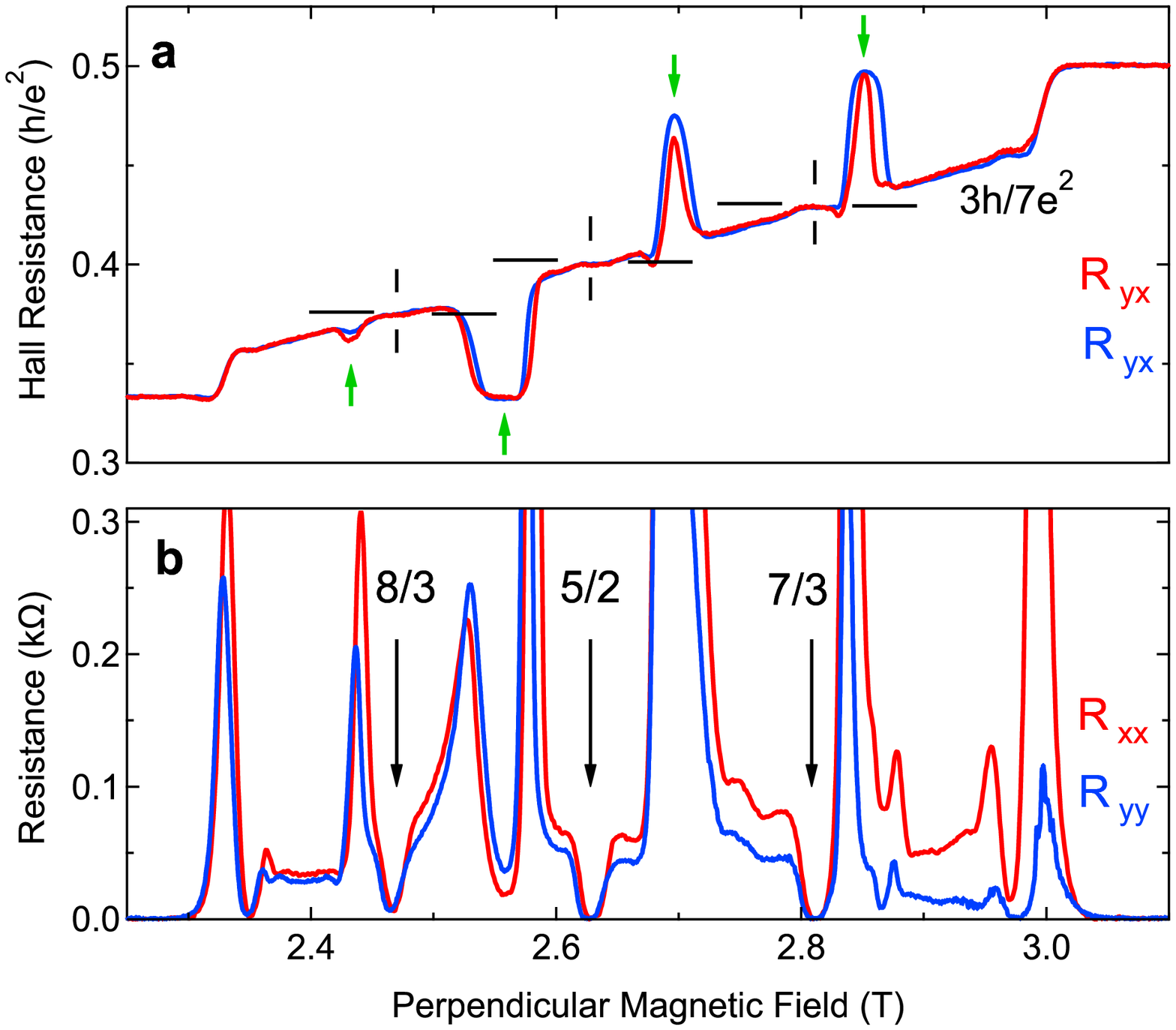}
\end{center}
\caption{Hall and longitudinal resistances at $T \approx 15$ mK vs. magnetic field in $N = 1$ Landau level, with $\nu = 7/3$, 5/2, and 8/3 FQHE states indicated by cross-hairs in a) and arrows in b).  \rxy\ and \rxx\ are the Hall and longitudinal resistances, respectively, for mean current flow along the \xdir\ direction; for \ryx\ and \ryy\ mean current flow is along \ydir. Sample is perpendicular to the magnetic field ($\theta = 0^\circ$).  Green arrows in a) indicate locations of re-entrant integer quantized Hall states.}
\label{fig1}
\end{figure}
Tilting the 2DES relative to the magnetic field direction has a profound influence on the various collective phases found in the $N = 1$ LL.  In agreement with prior studies \cite{JPE88,JPE90,JPE02,Csathy05}, we find that the $\nu = 5/2$ FQHE state and the re-entrant integer quantized Hall states are suppressed by an in-plane magnetic field component, \Bpar.  In addition to the destruction of these quantized Hall states, the general trend of the longitudinal resistance throughout the $N=1$ LL is to become increasingly anisotropic as \Bpar\ is initially applied, with \rxx\ (for which the mean current direction lies along \Bpar) growing significantly larger than \ryy\ \cite{Pan99,Lilly99b}.  

Figure 2 displays the temperature dependence of the longitudinal resistances \rxx\ and \ryy\ at $\nu = 7/3$ for various tilt angles $\theta$.  As expected, at $\theta = 0$ (Fig. 2a) we find \rxx\ and \ryy\ are very nearly equal at all temperatures.  Below about $T = 100$ mK the $\nu = 7/3$ FQHE begins to develop, with \rxx\ and \ryy\ dropping rapidly toward zero in unison as the temperature falls.  This temperature dependence is well-approximated by simple thermal activation, $R \sim {\rm exp}(-\Delta/2T)$, with $\Delta \approx 225$ mK.

\begin{figure}
\begin{center}
\includegraphics[width=1.0 \columnwidth] {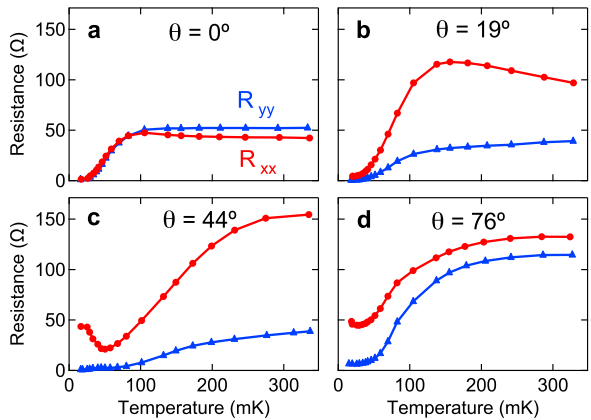}
\end{center}
\caption{\rxx\ (red dots) and \ryy\ (blue triangles) vs. temperature at $\nu = 7/3$ for $\theta = 0^\circ$, $19^\circ$, $44^\circ$, and $76^\circ$.}
\label{fig2}
\end{figure}
Tilting the sample to just $\theta = 19^\circ$ (Fig. 2b), where $B_{||}= 0.97$ T, creates a substantial anisotropy in the longitudinal resistance, with \rxx\ exceeding \ryy.  The anisotropy is present at relatively high temperatures ($T \sim 300$ mK), well above where both resistances begin to fall sharply as the FQHE develops.  Increasing the tilt angle to $\theta = 44^\circ$ (Fig. 2c), where $B_{||}=2.72$ T, enhances both the anisotropy and the temperature below which the resistances begin their FQHE-induced fall.  This latter effect reflects the tilt-induced increase of the $\nu = 7/3$ FQHE energy gap noted previously in similar samples \cite{Dean08,Xia10}.  At the large tilt angle of $\theta = 76^\circ$ (Fig. 2d), the anisotropy in the longitudinal resistance has subsided significantly at high temperatures, in spite of the large in-plane magnetic field ($B_{||}=11.3$ T) which breaks rotational symmetry.  This surprising return toward isotropic transport has been noted previously and related to mixing of the Landau levels emanating from the two lowest subbands of the confinement potential \cite{Xia10}.  

The $\theta = 44^\circ$ data in Fig. 2c reveal a curious effect: After falling steadily from $T \approx 250$ mK down to about 50 mK, \rxx, the resistance in the \xdir\ direction, suddenly begins to {\it rise} as the temperature is lowered further. No such anomaly is observed in \ryy, the resistance in the \ydir\ direction, which drops to very small values in the low temperature limit.  While Fig. 2d demonstrates that this peculiar behavior has almost vanished by $\theta = 76^\circ$, Fig. 3a shows it to be quite pronounced at $\theta = 66^\circ$, where \Bpar\ = 6.33 T at $\nu = 7/3$.

\begin{figure}
\begin{center}
\includegraphics[width=1.0 \columnwidth] {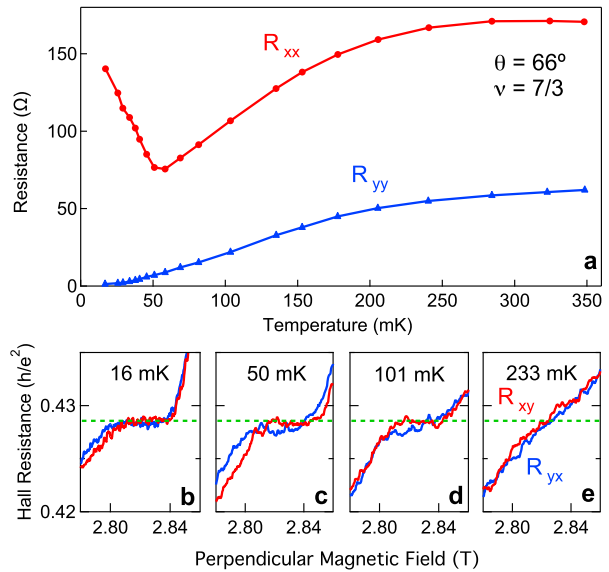}
\end{center}
\caption{Hall and longitudinal resistances at $\nu = 7/3$ for $\theta = 66^\circ$. a) \rxx\ and \ryy\ vs. temperature; b)-e) Hall resistances \rxy\ and \ryx\ vs. magnetic field at various temperatures. Green dashed line indicates expected location of $\nu = 7/3$ Hall plateau.}
\label{fig3}
\end{figure}
The $\theta = 66^\circ$ data in Fig. 3 reveal three distinct regimes of resistive anisotropy. In the high temperature regime ($350 > T>250$ mK) the resistances \rxx\ and \ryy\ are only weakly temperature dependent, with \rxx\ exceeding \ryy\ by about a factor of 3.  As Fig. 3e shows, the Hall resistances \rxy\ and \ryx\ do not exhibit a quantized Hall plateau in this regime.  In the intermediate temperature range $50<T<250$ mK, the FQHE is beginning to dominate the transport.  Both \rxx\ and \ryy\ fall with decreasing temperature and display local minima vs. magnetic field at $\nu = 7/3$ (see the supplementary information). As Fig. 3d illustrates, a Hall plateau at $R_{xy}=R_{yx}=3h/7e^2$ appears below about 150 mK.  The resistive anisotropy remains, with the ratio $R_{xx}/R_{yy}$ growing as the temperature falls.  (This behavior is inconsistent with simple thermal activation; see the supplementary information.)  Finally, there is the low temperature regime, demarcated by a sudden change in the sign of $dR_{xx}/dT$ at $T\approx 50$ mK.  Below this temperature \rxx\ rises steadily, ultimately reaching $R_{xx} \approx 150$ $\Omega$ at $T \approx 15$ mK, our lowest measurement temperature.  Throughout this low temperature range the Hall resistances \rxy\ and \ryx\ display well-quantized Hall plateaus, as Figs. 3b and 3c prove. Intriguingly, \ryy\ continues to fall toward zero in this low temperature regime, passing smoothly through $T \approx 50$ mK with no sign of the abrupt behavior exhibited by \rxx.

Consistent with prior work \cite{Pan99,Lilly99b,Cooper02}, when \Bpar\ is directed along \ydir\ instead of \xdir, the ``hard'' and ``easy'' axes of the low temperature resistive anisotropy at 7/3 switch to \ydir\ and \xdir, respectively.  Although the anisotropy is reduced in magnitude, the same curious upturn in the hard axis resistance (now \ryy) is observed at very low temperature.  At $\nu = 8/3$ we find that \Bpar\ induces a similarly strong resistive anisotropy without destroying the quantized Hall plateau, although no clear upturn in the hard axis resistance has so far been seen.

The resistive anisotropy for $T \gtrsim 250$ mK shown in Fig. 3 recalls that observed in the stripe phases of the $N\ge2$ LLs under similar tilted field conditions. Previous experiments on the $\nu = 9/2$ state have demonstrated that a weakly temperature dependent resistive anisotropy extends to high temperatures ($T \sim 500$ mK) when a strong in-plane magnetic field is present \cite{Cooper02}.  Furthermore, in common with the $\nu = 7/3$ anisotropy reported here, the orientation of this anisotropy is dictated by the direction of the in-plane magnetic field, with the high resistance direction usually parallel to \Bpar\ \cite{Pan99,Lilly99b,Dean08,Xia10,jungwirth,stanescu}. We therefore speculate that in the present highly tilted $\nu = 7/3$ case, some form of stripe-like density modulation is present at relatively high temperatures and is responsible for the observed anisotropy in the longitudinal resistance.  As the temperature is reduced, the FQHE begins to develop and compete with this stripe-like order.  Since the resistive anisotropy persists to considerably higher temperatures than the FQHE, the energy scale for the stripe-like state exceeds that of the FQHE state.  As a result, the quantum Hall fluid, which at these relatively high temperatures remains compressible, will accomodate itself to the density modulation.  Unless the modulation is too strong, we believe that it will simply be translated into a spatial modulation of the density of fractionally-charged quasiparticles above the FQHE gap.  The longitudinal resistance in such a weakly density-modulated FQHE state will certainly be anisotropic at finite temperatures and a quantized Hall plateau can be expected.  We note stripes only appear in the $N=1$ LL when a tilted field is applied, in contrast to the situation in the $N\ge2$ LLs. Quantum fluctuations \cite{FK} are much stronger in the $N=1$ LL and it seems plausible that they will limit the density modulation in any tilt-induced stripe phase.

While the above scenario may explain the data in Fig. 3 in the high and intermediate temperature ranges, it does not readily account for the upturn in \rxx\ at $T \approx 50$ mK and the persistence of the Hall plateau down to $T \approx 15$ mK.  Indeed, the abruptness of the upturn suggests the emergence of a new electronic configuration in the low temperature regime.  The $\nu = 7/3$ longitudinal resistance data shown in Fig. 3a are, for $T \lesssim 50$ mK, again reminiscent of that seen in the stripe phases at half filling of the $N \ge 2$ LLs (e.g. at $\nu = 9/2$, 11/2, etc.) only now as they appear in the {\it absence} of an in-plane magnetic field.  Under these conditions a strong resistive anisotropy develops only at low temperatures ($T\lesssim 100$ mK). The resistance becomes large in one crystallographic direction and very small in the orthogonal direction.  Here, at $\nu = 7/3$, \rxx\ rises rapidly below $T = 50$ mK, reaching approximately 150 $\Omega$ at $T \approx 15$ mK, while \ryy\ concurrently falls to $\sim 1~\Omega$.  However, in sharp contrast to the anisotropic phases at $\nu = 9/2$, 11/2, etc., the anisotropic $\nu = 7/3$ state exhibits a robustly quantized Hall plateau down to the lowest temperatures studied.  

The simultaneous presence, at $\nu = 7/3$ in the present instance, of an accurately quantized Hall plateau and strong, highly temperature dependent anisotropy in the longitudinal resistance, has not been encountered experimentally before. There have however, been theoretical suggestions of fractional quantized Hall phases with broken rotational symmetry.  Using a variational approach, Musaelian and Joynt (MJ) \cite{Joynt96} suggested that in wide quantum wells (which soften the short-range part of the Coulomb interaction) the 2DES at $\nu = 1/3$ in the $N=0$ Landau level might be most stable in a phase analogous to a classical nematic liquid crystal. While exhibiting gapless neutral collective modes of the director order parameter, MJ nonetheless argue that charged excitations are gapped and thus the system would still exhibit the FQHE.  How the transport coefficients would behave as functions of temperature, and what might pin the order parameter, was not addressed.  More recently, Mulligan, Nayak, and Kachru (MNK) \cite{Nayak}, using an effective field theory approach, have predicted a transition from an isotropic to an anisotropic FQHE state with nematic order.  The transition is driven by subtle changes in the electron-electron interaction which, in the light of the present experimental results, MNK speculated arise from couplings between an in-plane magnetic field and the finite thickness of the 2DES.  While MNK predict that both \rxx\ and \ryy\ will ultimately vanish at $T = 0$, they also find a regime, as in our experiments, where one of the two resistances increases as the temperature falls while the other falls.  

In addition to the homogeneous nematic FQHE phases mentioned above, it is also possible that our results reflect a phase separated 2DES. In analogy with the situation at $\nu = 9/2$, a simple stripe phase picture at $\nu = 7/3$ (consisting, for example, of alternating stripes of $\nu = 2$ and $\nu = 3$) would yield anisotropic longitudinal transport but not quantization of the Hall resistance.  (We are ignoring the possible formation of isotropic ``bubble'' phases \cite{KFS,FKS,MC} away from half filling.) Alternatively, one can imagine that at $\nu = 7/3$ the electrons in the 1/3-filled $N=1$ Landau level exist in an anisotropic version of the quantized Hall insulator (QHI) first encountered experimentally at $\nu \lesssim 1/3$ in the $N=0$ LL in relatively disordered samples \cite{Shimshoni,Hilke}.  In Shimshoni and Auerbach's (SA) theory \cite{Shimshoni}, the QHI is modeled as a collection of incompressible FQHE puddles immersed in an insulating background fluid and connected to one another by tunnel junctions. SA find that the longitudinal resistance diverges as $T \rightarrow 0$ and, remarkably, the Hall resistance is quantized at the value appropriate to the FQHE puddles. Adjusting this scenario to $\nu \lesssim 7/3$=2+1/3 suggests a Hall resistance quantized at $R_{xy} = 3h/7e^2$ and a longitudinal resistance $R_{xx}$ which, though vanishing at $T=0$, exhibits quasi-insulating behavior ($dR_{xx}/dT<0$) in some intermediate temperature range. By construction, the SA model is isotropic.  Thus, while our \rxx, \rxy, and \ryx\ data are broadly consistent with the SA model, it does not encompass the behavior we observe in \ryy. Nevertheless, it seems at least plausible that by assuming the FQHE puddles to be oblate and consistently oriented relative to the in-plane magnetic field, that anisotropy in the longitudinal resistance would emerge from the model. Whether the different temperature dependences of \rxx\ and \ryy\ can also be accomodated is less clear.

{\bf Methods}

The sample we employ is a standard GaAs/AlGaAs heterostructure grown by molecular beam epitaxy.  A 40 nm GaAs quantum well is embedded in thick Al$_{0.24}$Ga$_{0.76}$As cladding layers.  Si doping sheets in each cladding layer create a 2DES in the lowest subband of the GaAs quantum well.  After illumination the density and mobility of this 2DES are $n = 1.6 \times 10^{11}$ cm$^{-2}$ and $\mu = 16 \times 10^6$ cm${^2}$/Vs, respectively.  The sample is a 5 mm square chip whose edges are parallel to the \xdir\ and \ydir\ crystal directions, henceforth referred to as the $\hat{x}$ and $\hat{y}$ directions, respectively.  InSn ohmic contacts are positioned at the corners and side midpoints of the sample.  Longitudinal resistance (\rxx\ and \ryy) measurements are performed by driving ac current (typically 2 nA at 13 Hz) between midpoint contacts on opposite sides the sample and detecting the resulting voltage difference between corner contacts on one side of the mean current axis.  For the Hall resistances (\rxy\ and \ryx), the voltage difference between the two midpoint contacts on opposite sides of the current axis is recorded.  The sample is mounted on a rotating platform (allowing for the application of an in-plane magnetic field component) thermally linked to the mixing chamber of a dilution refrigerator.

{\bf Acknowledgements}

We are grateful to Chetan Nayak and Steve Kivelson for useful discussions.  This work was supported by Microsoft Project Q. The work at Princeton was partially funded by the Gordon and Betty Moore Foundation as well as the National Science Foundation MRSEC Program through the Princeton Center for Complex Materials (DMR-0819860).

{\bf Author contributions}

J.X and J.P.E conceived the project.  L.N.P and K.W.W fabricated the samples. J.X performed the experiment. J.X and J.P.E discussed the data and co-wrote the manuscript.

{\bf Additional information}
The authors declare no competing financial interests.

\end{document}


\title{Evidence for a fractionally quantized Hall state with anisotropic longitudinal transport \\* (Supplementary information)}

\author{Jing Xia$^1$, J.P. Eisenstein$^1$, L.N. Pfeiffer$^2$, and K.W. West$^2$}

\affiliation{$^1$Condensed Matter Physics, California Institute of Technology, Pasadena, CA 91125
\\
$^2$Department of Electrical Engineering, Princeton University, Princeton, NJ 08544}

\date{\today}



\pacs{73.43.-f, 73.43.Nq, 73.21.Fg}

\maketitle

The energy gap $\Delta$ of a fractional quantized Hall effect (FQHE) state is most often determined via the temperature dependence of the longitudinal resistance, $R$.  Generally, there is some range of temperature over which $R \sim$ exp$(-\Delta/2T)$.  Since ordinary FQHE states are isotropic in the 2D plane, one expects the value of the gap $\Delta$ to be independent of whether \rxx\ or \ryy\ is used for its determination.  Figure S1 a) shows that this is very nearly the case for the $\nu = 7/3$ FQHE state in our sample when the magnetic field is perpendicular to the plane (tilt angle $\theta = 0$). The simple linear fits to log($R$) $vs.$ $T^{-1}$ shown in the figure give gaps of 240 mK and 210 mK for the \rxx\ and \ryy\ data, respectively.  We do not believe that the small difference between these values is statistically significant and so give $\Delta \approx 225$ mK as our best estimate of the $\nu = 7/3$ gap at $\theta = 0$.

As our paper reports, tilting the sample relative to the magnetic field induces substantial anisotropy in the longitudinal resistance at $\nu = 7/3$ even though a quantized Hall plateau remains.  We find, however, that for tilt angles beyond $\theta \approx 20^\circ$, it is no longer possible to extract a unique energy gap from the temperature dependence of \rxx\ and \ryy.  This difficulty is illustrated in Fig. S1 b) where log($R_{xx}$) and log($R_{yy}$) at $\nu = 7/3$ and $\theta = 66^\circ$ are plotted $vs.$ $T^{-1}$. Even leaving aside the dramatic divergence of these two resistances for $T<50$ mK, it is clear from the figure that in the intermediate temperature range $50<T<250$ mK their temperature dependences are significantly different and only weakly resemble simple thermal activation. Figure S1 c) shows the results of separately fitting \rxx\ and \ryy\ at $\nu = 7/3$ in the intermediate temperature range to the standard Arrhenius form exp$(-\Delta/2T)$. The extracted values of $\Delta_x$ and $\Delta_y$ (from the \rxx\ and \ryy\ data, respectively) are plotted versus the in-plane magnetic field \Bpar\ resulting from tilting the sample by an angle $\theta$.  As reported previously \cite{Dean08,Xia10}, the energy gap for the $\nu = 7/3$ FQHE state in similar samples initially increases with \Bpar.  However, beyond about $\theta = 20^\circ$, $\Delta_x$ and $\Delta_y$ become significantly different, rendering a simple interpretation of them as energy gaps for the FQHE state suspect. 

Figure S1 c) demonstrates that the \Bpar\ dependence of $\Delta_x$ and $\Delta_y$ at $\nu = 7/3$ is quite complex and not readily interpreted.  It is remarkable that in spite of this complexity a robust Hall plateau at $R_{xy}=R_{yx}=3h/7e^2$ is observed at low temperatures at all tilt angles we have studied ($0\le \theta \le 76^\circ$).  

Figure S2 shows these Hall plateaus, along with the associated local minima in \rxx\ and \ryy\ at several tilt angles.

\begin{figure}
\begin{center}
\includegraphics[width=0.6 \columnwidth] {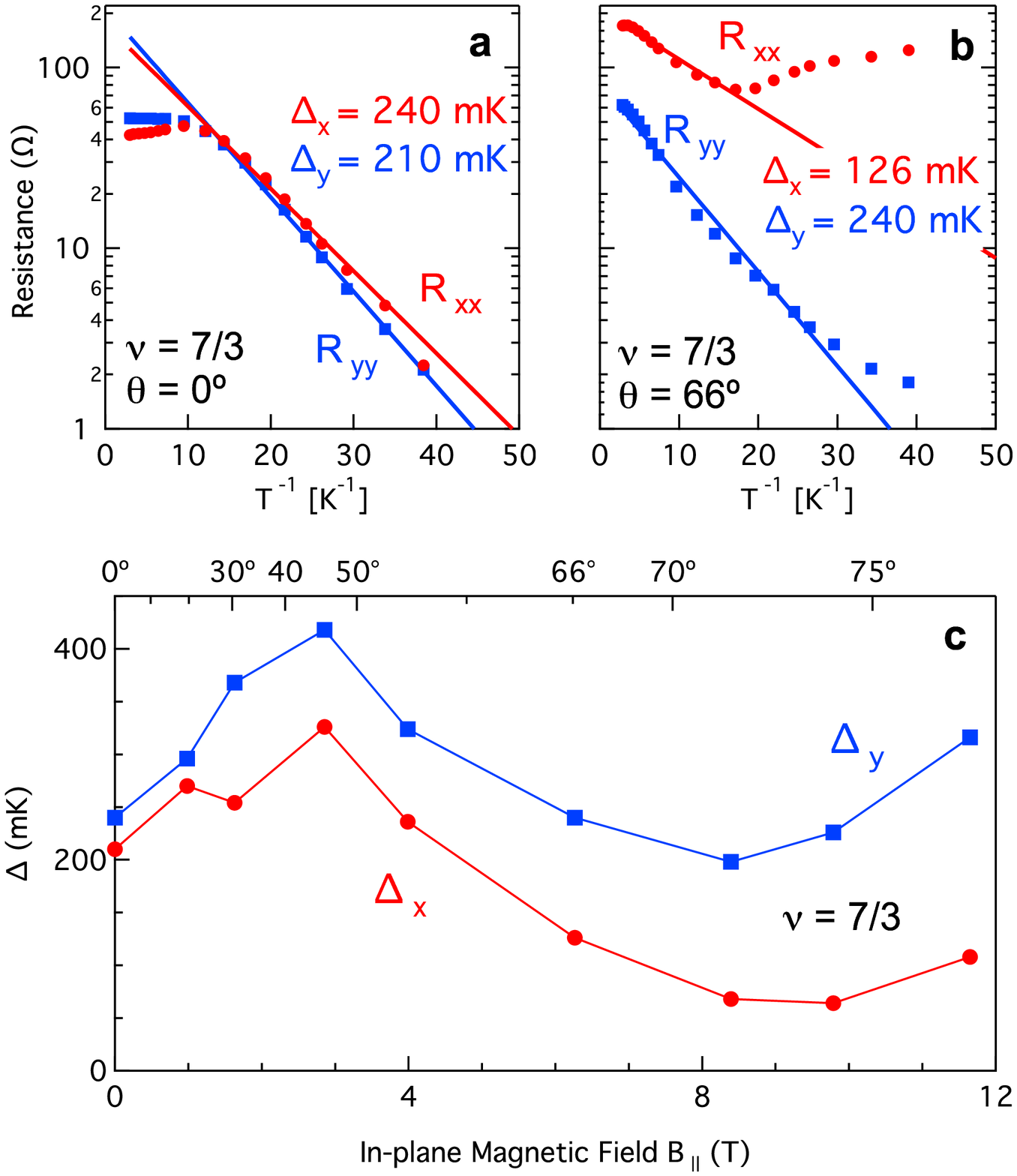}
\end{center}
\caption{a) and b) Arrhenius plots of \rxx\ and \ryy\ at $\nu=7/3$ $vs.$ $T^{-1}$ at $\theta = 0^\circ$ and $66^\circ$, respectively.  $\Delta_x$ and $\Delta_y$ are the gap parameters extracted from fitting the data in the intermediate temperature range $50<T<250$ mK. c)  Fitted gap parameters $\Delta_x$ and $\Delta_y$ $vs.$ in-plane magnetic field \Bpar\ and tilt angle $\theta$. }
\label{fig1}
\end{figure}

\begin{figure}
\begin{center}
\includegraphics[width=1.0 \columnwidth] {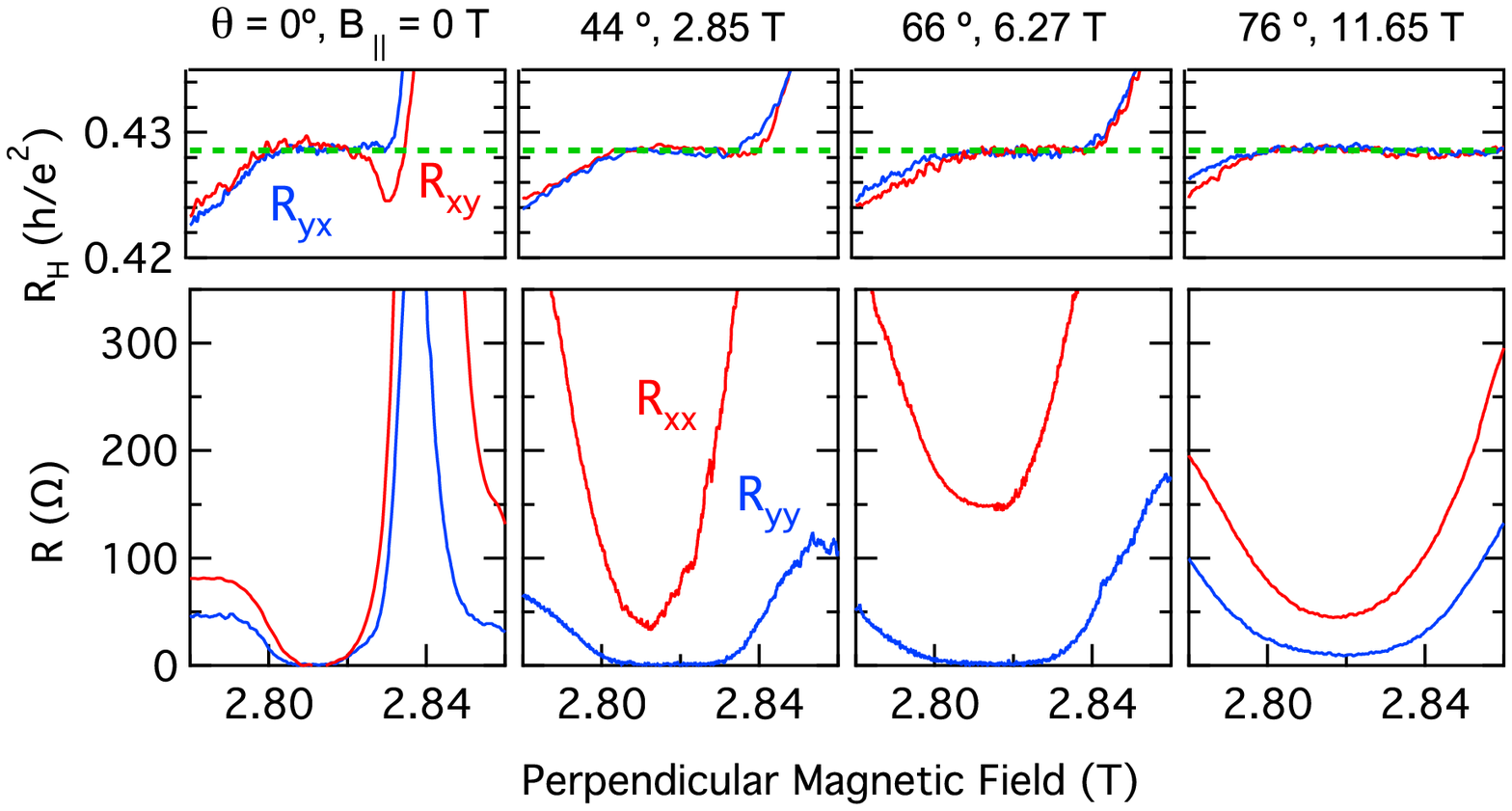}
\end{center}
\caption{Hall plateaus and longitudinal resistance minima at $\nu = 7/3$ for various tilt angles and in-plane magnetic fields. All data at $T \approx 15$ mK.}
\label{fig2}
\end{figure}